\documentclass[pamm,a4paper,fleqn]{w-art}
\usepackage{times,cite,w-thm}
\usepackage[T1]{fontenc}
\usepackage[utf8]{inputenc}
\usepackage{graphicx}
\usepackage{amsmath}
\usepackage{amssymb}
\usepackage{booktabs}
\usepackage{url}

\usepackage{todonotes}
\usepackage{xcolor}
\usepackage{tikz}
\usetikzlibrary{shapes,positioning,shadings}
\usepackage{pgfplots}
\pgfplotsset{compat=1.18} 

\begin{document}

\def\leftmark{G. Ramirez-Hidalgo et al.: Data-Driven Spectral Prediction for Electronic Structure Calculations}
\def\rightmark{Preprint}

\TitleLanguage[EN]
\title[Data-Driven Spectral Prediction]{Data-Driven Spectral Prediction for Accelerating Large-Scale Electronic Structure Calculations}

\author{\firstname{Abhiram} \lastname{Badrinarayanan}\inst{2}}
\author{\firstname{Davor} \lastname{Davidovi\'{c}}\inst{2}}
\author{\firstname{Edoardo} \lastname{Di Napoli}\inst{1}}
\author{\firstname{Jurica} \lastname{Novak}\inst{2}}
\author{\firstname{Luigi} \lastname{Genovese}\inst{3}}
\author{\firstname{Gustavo} \lastname{Ramirez-Hidalgo}\inst{1,}%
\footnote{Corresponding author: e-mail \ElectronicMail{g.ramirez.hidalgo@fz-juelich.de}}} 

\author{\firstname{Xinzhe} \lastname{Wu}\inst{1}}

\address[\inst{1}]{\CountryCode[DE]Jülich Supercomputing Centre, Forschungszentrum Jülich, Germany}
\address[\inst{2}]{\CountryCode[HR]Ruđer Bošković Institute, Croatia}
\address[\inst{3}]{\CountryCode[FR]CEA / IRIG / MEM / L\_Sim, France}

\AbstractLanguage[EN]
\begin{abstract}
Simulating large molecular systems comprising thousands of atoms requires highly scalable methodologies. While modern Density Functional Theory (DFT) codes exhibit linear scaling, solving the associated large, sparse generalized eigenproblems remains a critical computational bottleneck on exascale architectures. In the context of the LimitX project, we propose a data-driven framework to accelerate these calculations. By shifting the machine learning target from discrete eigenvalues to the coefficients of an interpolating Chebyshev polynomial, and by comparing both all-atom and fragment-based structural representations, we successfully overcome the dimensionality constraints of large-scale spectral prediction. We investigate three machine learning models (Kernel Ridge Regression, Graph Neural Networks, and Random Forests) trained on a novel 2 TB dataset of protein dimers. The predicted spectra provide initial guesses that effectively bypass early Self-Consistent Field (SCF) iterations in BigDFT. Ultimately, these spectral predictors will be deployed to dynamically optimize upcoming rational filter-based eigensolvers, such as FrASE, which is currently in initial development.
\end{abstract}
\maketitle
\thispagestyle{empty}

\section{Introduction}

In contemporary scientific exploration within the realms of solid-state physics, materials science, chemistry, and biology, there exists a pressing need to simulate large molecular systems comprising thousands of atoms. Density Functional Theory (DFT) calculations have emerged as the most promising methodology to tackle these physical systems in terms of both scalability and computational performance. A notable example is the BigDFT code, which utilizes Daubechies wavelets as a basis set \cite{mohr2014daubechies}. This approach demonstrates exceptional parallel performance -- exhibiting highly efficient weak scaling as well as strong scaling efficiencies exceeding 90\% when measured relative to a multi-core baseline node -- enabling the capacity to routinely analyze highly complex biomolecular and soft-matter systems. Similarly, complementary approaches that integrate large-scale first-principle DFT codes with interior eigenvalue solvers, such as Conquest, have achieved notable advancements in accuracy and efficiency for systems surpassing 10,000 atoms \cite{nakata2017conquest}.

Despite these algorithmic strides achieving linear scaling with respect to the number of atoms, solving the underlying large and sparse numerical linear systems and generalized eigenproblems remains the most computationally expensive portion of the DFT workflow. These numerical linear algebra routines constitute the computational pillars for numerous large-scale scientific applications. Ensuring their scalability and efficiency is absolutely crucial before embarking on higher-level application software optimization, especially as the community transitions toward massively parallel exascale computing architectures where load balancing becomes a primary constraint.


To address the delicate balance between accuracy and computational efficiency, research has increasingly turned toward adaptive recommender systems to optimize solver selection. Early methodologies from the mid-1990s adopted an analytical approach, encompassing poly-iterative solvers (methods that simultaneously execute multiple iterative algorithms), composite multi-method solvers, and adaptive Newton-Krylov algorithms that evaluated and selected appropriate solvers during the simulation stages \cite{barrett1996polyiterative, bhowmick2002composite}. More recently, the community has framed this as a machine learning classification problem. For instance, models have been trained on structural attributes (which describe the matrix sparsity structure), norm-related properties, and the statistical variance of matrix entries to categorize matrices as "suitable" or "unsuitable" for specific solvers \cite{bhowmick2006application}. Sophisticated feature extraction techniques—including eigenvalue estimates—have yielded exact classification accuracy rates of up to 87\% \cite{sood2015automated}, while other efforts have sought to predict ideal solvers using Convolutional Neural Networks (CNNs) based on computationally cheaper high-level features \cite{funk2022prediction}. Alternative paradigms have even explored Physics-Informed Neural Networks (PINNs) hybridized with traditional solvers like Gauss-Seidel or multi-grid methods \cite{markidis2021pinns}.


In this work, developed within the context of the LimitX (Learning Materials at eXascale) project, we aim to explore a third path. We want to maintain a high accuracy of our prediction but, at the same time, restrict the field of applicability of our methodology. In an environment where "generalization" of AI models seems to be the paramount objective, we claim that the chemical surrogate space is too vast to be generalized and prediction should be contained within the specific area of validity dictated by the physical phenomena it describes. Such an approach implies that, while our methods are general and "generalizable", the model we generate by training on a specific set of data is not. The novelty of our contribution resides in this dichotomy: on the one side we use a number of different learning models ranging from traditional to the innovative. On the other hand, we use our model on a class of problems that is very large (from the Quantum Physics perspective) but limited to a class of biological molecules with affine properties. In practice, by treating sparse matrices akin to image recognition tasks and avoiding computationally prohibitive spectral feature extraction, we develop a recommender system that forecasts spectral properties directly. The highlight of this approach is that we actually develop not one but three spectral predictors which all agree when tested on the available data. Integrating this predictor with a highly curated database of matrices representative of the Quantum Physics surrogate space allows us to accurately fine-tune preconditioners. These preconditioners are then used to significantly accelerate the solution of very large linear systems from which is possible to extract the eigenvectors corresponding to the predicted spectrum. Not only such a method would accelerate the simulation of atomistic and molecular systems, but it would also do so by maximally exploiting the parallelism of current and upcoming supercomputing clusters.

\section{Data Generation, Representation and Reduction}

\subsection{Data set}
Machine learning models require large amounts of high-quality data. To this end, we generated a comprehensive dataset of Hamiltonian and overlap matrices derived from full Linear Scaling (LS) BigDFT simulations of 423 different protein--protein complexes (dimers). In the end, we assembled a curated descriptor archive of 423 complexes spanning 687--16,260 atoms (median 5,908), covering the intended range from approximately 1,000 to over 15,000 atoms. For each dimer, the reference eigenvalue spectrum was obtained by solving the generalized symmetric eigenproblem associated with the BigDFT Hamiltonian and overlap matrices using SLEPc~\cite{hernandez2005slepc}, requesting the algebraically smallest real eigenvalues (\texttt{eps\_smallest\_real}).

All simulations were performed with BigDFT v1.9.4 in linear-scaling mode using a real-space grid spacing of $h_{\mathrm{grid}} = 0.45$~bohr, and an amino-acid fragment neighbour cutoff of 12.0~\AA. The resulting LS basis contains 911--23,065 Kohn--Sham orbitals. Table~\ref{tab:dataset_stats} summarizes the principal size statistics of the processed archive.
\begin{table}[t]
\centering
\caption{Summary statistics of the LimitX protein-dimer descriptor archive (423 processed systems).}
\label{tab:dataset_stats}
\begin{tabular}{lrrrrr}
\toprule
Quantity & Min & Q1 & Median & Q3 & Max \\
\midrule
Atoms                          &    687 &  3,654 &  5,908 &  9,198 & 16,260 \\
Amino acid residues            &     42 &    240 &    377 &    585 &  1,095 \\
Protein chains                 &      2 &      2 &      2 &      4 &      9 \\
Fragments (12.0~\AA{} cutoff)  &     30 &    210 &    319 &    432 &  1,009 \\
Kohn--Sham orbitals            &    911 &  5,156 &  8,237 & 12,786 & 23,065 \\
Support functions              &  1,647 &  9,215 & 14,810 & 22,971 & 41,232 \\
Electrons                      &  1,822 & 10,313 & 16,474 & 25,571 & 46,130 \\
\bottomrule
\end{tabular}
\end{table}
The majority of systems (53\%) contain between 3,000 and 9,000 atoms, with fewer complexes at the lower ($<$\,3,000 atoms; 17\%) and upper ($>$\,9,000 atoms; 27\%) ends of the size distribution. The smallest processed dimer is \texttt{1U0I} (687 atoms, 30 fragments) and the largest is \texttt{1ZA3} (16,260 atoms, 523 fragments).

A primary challenge in applying machine learning to such large molecules is the sheer scale of the input features. To address this, we compared traditional all-atom representations against a fragment-based abstraction. For the fragment-based approach (utilized by KRR and GNN), we abstracted the physical systems using amino-acid fragments automatically detected by BigDFT. Each dimer is represented by a sparse adjacency matrix derived from fragment bond-order/contact information, with inter-fragment neighbourhoods evaluated at multiple cutoffs (11.0--15.0~\AA{} in steps of 0.25~\AA), rather than by storing the full set of atomic coordinates as input features. At the default 12.0~\AA{} cutoff, this reduces the input from ${\sim}6,500$ atoms on average to ${\sim}340$ fragment nodes (${\sim}20\times$ compression), while preserving the essential structural and spatial relationships necessary for accurate spectral prediction. Conversely, for the Random Forest models, we utilized an all-atom representation encoded via Radial Distribution Functions (RDF) as a comparative baseline.
To overcome the complementary challenge of variable-length spectral outputs, with converged eigenvalue spectra range from ${\sim}3,000$ to ${\sim}330,000$ values depending on system size, each spectrum is also represented by a fixed vector of 10 Chebyshev expansion coefficients (\texttt{cheb\_coeffs\_10}), providing a size-independent regression target for KRR and a reference reconstruction for all models (see Section~\ref{subsection: Overcoming Dimensionality}).

\begin{figure}[htbp]
    \centering
    \includegraphics[width=0.95\textwidth]{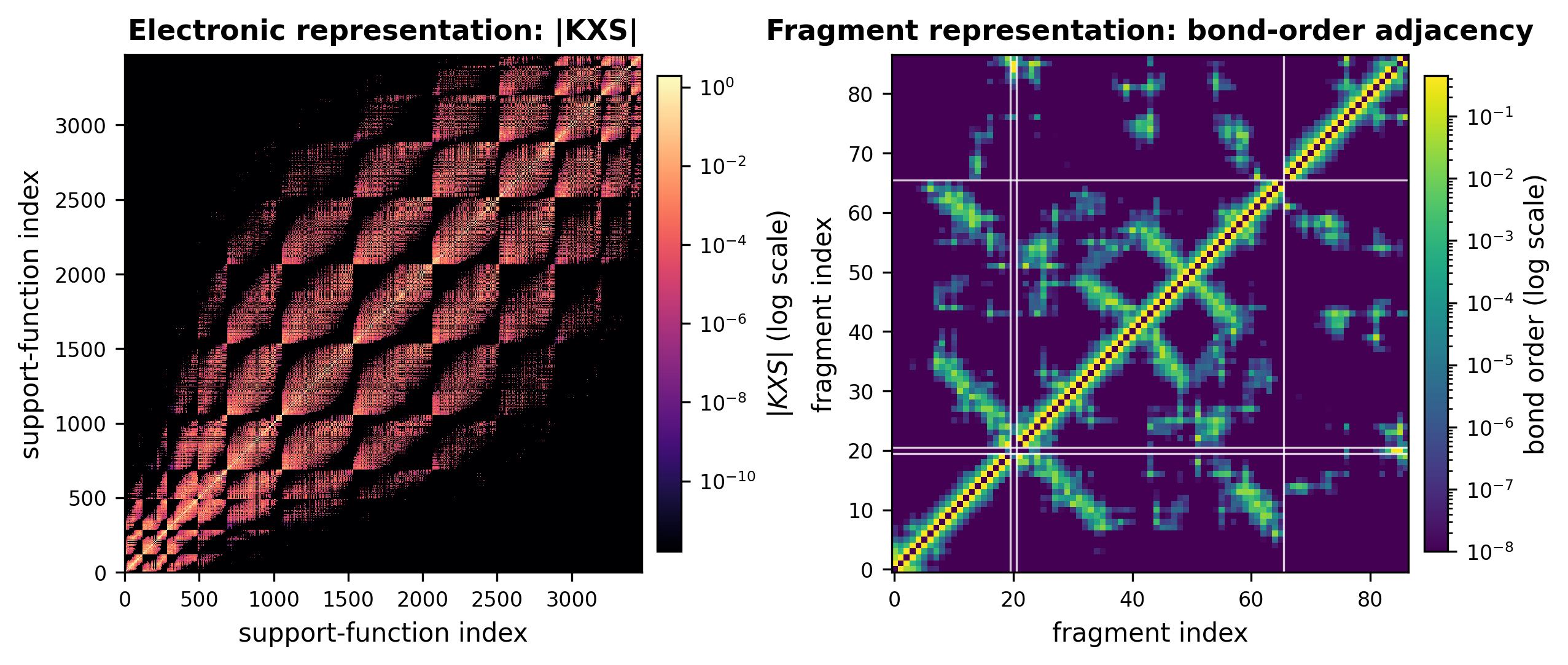}
    \caption{Comparison between the electronic matrix representation and the fragment-level representation for the 1ZSG system. Left: logarithmic absolute-value view of the sparse $KXS$ matrix in the support-function basis. Right: fragment bond-order adjacency matrix extracted from the serialization data, with diagonal self-contributions removed to emphasize inter-fragment connectivity. The fragment representation compresses the original electronic problem into a chemically meaningful graph whose nodes are amino-acid fragments and whose edges encode bond-order/contact strength.}
    \label{fig:kernel_bond_order_representation}
\end{figure}

Figure~\ref{fig:kernel_bond_order_representation} illustrates the reduction used in the learning pipeline. The support-function matrix retains the electronic degrees of freedom needed by the underlying DFT calculation, whereas the fragment bond-order adjacency matrix provides a much smaller graph representation that still reflects the chemically relevant connectivity of the protein complex. This distinction is important for the present work: the learning models are not trained on the full sparse electronic matrices directly, but on compact descriptors derived from fragment-level chemical and geometrical information.




\subsection{Overcoming Dimensionality}\label{subsection: Overcoming Dimensionality}

Predicting the full eigenspectrum of a large molecular system is computationally prohibitive. Direct prediction requires neural networks to output vectors containing tens or hundreds of thousands of individual eigenvalues. Furthermore, the length of this output vector varies drastically depending on the size of the input system, complicating the architecture and training of standard machine learning models.

To overcome this dimensionality challenge, we switched to a simpler target. Instead of predicting discrete eigenvalues, our models are trained to predict the expansion coefficients of a Chebyshev polynomial that interpolates the eigenspectrum. By expressing the spectrum as an expansion of Chebyshev polynomials of the first kind, $T_k(x)$, the predicted continuous spectrum $p(x)$ is reconstructed as:
\begin{equation}
    p(x) = \sum_{k=0}^{N} c_k T_k(x)
\end{equation}
This approach successfully transforms the highly variable, massive-scale prediction problem into the prediction of a fixed-length vector of Chebyshev coefficients ($c_k$).

Alternatively, the spectral estimation problem can be framed as predicting the histogram bin occupancy, effectively approximating the spectral density of states. The underlying eigenspectrum is discretised into a predefined number of bins $N$ with uniform width $\Delta = \frac{\lambda_{\max} - \lambda_{\min}}{N}$. The model is then trained to predict the frequency vector $\mathbf{y} = [y_1, y_2, \dots, y_N]^T \in \mathbb{N}_0^N$ of eigenvalues, where each component $y_i$ represents the frequency or count of eigenvalues within the bin $i$. To reconstruct the continuous spectrum from these discrete predictions, we apply linear interpolation within each bin and sort obtained eigenvalues.
These two approaches ensure a scalable, efficient, and generalizable learning process regardless of the molecular system's size.

\section{Theoretical Foundation of Predictive Machine Learning Models}
To map the structural representations of the molecular systems to the target Chebyshev coefficients, we investigated three distinct machine learning paradigms. Each method offers a different balance between computational complexity, interpretability, and predictive capacity.

\paragraph{Kernel Ridge Regression (KRR):} 
Kernel Ridge Regression is a highly interpretable, non-parametric method that combines ridge regression ($L_2$ regularization) with the kernel trick, allowing for the modeling of complex, non-linear relationships by implicitly mapping input features into a high-dimensional vector space \cite{murphy2012machine}. In the context of computational chemistry, KRR has been widely successful when paired with global molecular descriptors, such as Coulomb matrices, which encode the electrostatic interactions between atomic nuclei \cite{rupp2012fast}. While Coulomb matrices provide a translationally and rotationally invariant representation, directly applying them to arbitrarily sized macro-molecules poses significant challenges for vector normalization. To circumvent this, spectral representations of these matrices are often compressed or randomized to yield fixed-size feature vectors suitable for the KRR kernel functions (e.g., linear, polynomial, or Radial Basis Functions).

\paragraph{Graph Neural Networks (GNN):} 
Graph Neural Networks have recently revolutionized molecular property prediction by directly operating on the topological structure of molecules \cite{gilmer2017neural}. In this framework, physical systems are modelled as graphs $\mathcal{G} = (\mathcal{V}, \mathcal{E})$, where nodes $\mathcal{V}$ represent atomic or fragment entities and edges $\mathcal{E}$ encode spatial proximities or chemical bonds. We leverage the continuous-filter convolutional neural network architecture, SchNet \cite{schutt2017schnet}, which is specifically designed to model quantum interactions. SchNet employs a message-passing scheme where the hidden state of each node is iteratively updated using aggregated information from its local neighborhood. This continuous-filter approach respects the rotational and translational invariances of the physical system, making it highly effective at capturing the complex, many-body interactions necessary for accurately predicting the overall shape of the eigenspectrum. While the standard SchNet architecture outputs a single scalar, in this work, we use a modified version of the SchNet architecture that outputs a vector corresponding to the predicted histogram bin occupancies of the converged spectrum.


\paragraph{Random Forests and Radial Distribution Functions (RDF):}
Random Forests (RF) are robust ensemble learning algorithms that construct a multitude of decision trees during training and output the mean prediction of the individual trees, effectively mitigating the risk of overfitting \cite{breiman2001random}. To apply RF to molecular spectral prediction, the input data must be transformed into a fixed-length vector. Molecular structures were represented using the weighted radial distribution function (RDF), a three-dimensional molecular descriptor introduced by Hemmer et al.~\cite{Hemmer1999}. The RDF encodes the spatial arrangement of atoms by accumulating pairwise contributions over all atom pairs $(i, j)$ according to
\begin{equation}
g(r) = f \sum_{i}^{N-1} \sum_{j>i}^{N} A_i A_j \, 
\exp\!\left(-B\left(r - r_{ij}\right)^2\right),
\label{eq:rdf}
\end{equation}

\noindent where $r$ is the interatomic distance, $r_{ij}$ the Euclidean distance between atoms $i$ and $j$, $A_i$ and $A_j$ atomic properties assigned to atoms $i$ and $j$ respectively, $B = 1/\Delta r^2$ a smoothing parameter dependent on the discretization step $\Delta r$, and $f$ a scaling factor~\cite{Hemmer1999}. The resulting descriptor vector $\mathbf{g} \in \mathbb{R}^{M}$, sampled over $M$ equidistant points in the range $r \in [0,\, 100~ \text{\AA}]$, constitutes a fixed-length numerical fingerprint of the molecular geometry. To assess the influence of atomic weighting on predictive performance, five choices of atomic property $A$ were evaluated: an unweighted variant ($A_i = A_j = 1$), number of valence electrons, covalent radius, van der Waals radius, and Pauling electronegativity. These RDF feature vectors were subsequently used as input to train the Random Forest models.

\section{Implementation Details and Results}

To validate our predictive approach, we established rigorous implementation and testing pipelines for the considered models. 

\subsection{Kernel Ridge Regression}

\paragraph{Data Preparation and Feature Engineering:}
The raw input format initially proposed for the model was the full distance (Coulomb) matrix describing the fragments in the physical system. However, this strategy was discarded due to the severe challenge of normalizing highly variable input sizes across diverse molecular systems. Consequently, our final feature engineering pipeline utilized a randomized compression on the real spectrum of these matrices. 

For each protein complex, we computed the dense array of the adjacency distance matrix, extracted its exact eigenvalues, and used them as input for regression. To form a standardized input vector $X$, we randomly sampled a fixed subset of up to 500 eigenvalues from the real spectrum of the associated adjacency distance matrix. If a system produced fewer than 500 eigenvalues, the vector was uniformly padded with zeros ($0$) to reach the correct dimensionality. Finally, the entire vector was normalized via its $L_2$ norm before being passed to the regressor.

In contrast, the target output $Y$ was constructed by the pre-computed Chebyshev coefficients, which interpolates the real eigenspectrum of the Hamiltonian of each system (Section \ref{subsection: Overcoming Dimensionality}). To maintain uniform dimensionality, the model was restricted to predicting exactly 10 coefficients per system, achieved by zero-padding sparse vectors or truncating longer ones.

\paragraph{Model Training and Hyperparameter Optimization:}
The KRR regressor was trained using the \texttt{scikit-learn} framework, utilizing a robust 5-fold cross-validation Grid Search to optimize hyperparameters. The exhaustive parameter grid investigated:
\begin{itemize}
    \item \textbf{Regularization strength ($\alpha$):} Evaluated at $0.0001$, $0.01$, and $20.0$.
    \item \textbf{Kernel functions:} Tested linear, polynomial, radial basis function (RBF), and sigmoid kernels.
    \item \textbf{Kernel-specific parameters:} Polynomial degrees ranging from 2 to 5, and RBF kernel coefficients ($\gamma$) evaluated at $0.0001$, $0.001$, and $0.01$.
\end{itemize}

A critical aspect of the training phase was the application of a custom scoring function. The standard mean squared error uniformly penalizes discrepancies across the spectrum. To strictly penalize proportional errors, the grid search utilized a custom negative weighted mean squared error evaluator, which explicitly applied the absolute values of the true targets as sample weights during the error calculation. The resulting model was evaluated across the test subset using an array of regression metrics: Mean Squared Error (MSE), Mean Absolute Error (MAE). 

\subsection{Graph Neural Networks}

\paragraph{Architecture and Node Embeddings:}
The GNN pipeline utilizes a modified Continuous-filter convolutional Graph Neural Network (SchNet) implemented via PyTorch Geometric \cite{schutt2017schnet}. The standard SchNet model is modified at the very last layer to produce a vector output instead of a scalar one. The vector elements correspond to the bin occupancies of the histogram of the predicted eigenspectrum. Thus, with this method, we effectively predict the density of states with a given resolution.

In standard atomistic networks, nodes are typically embedded using atomic numbers. To adapt the network for our case, we explored two different embeddings. In the first case, nodes - each representing a fragment (i.e. amino-acids), are assigned a unique diagonal value of the fragment adjacency matrix. This embedding considers the variations in the hydrogen atom content of the fragment, meaning that two nodes that correspond to the same amino acid but have a different number of hydrogen atoms will have different embeddings. In the second case, the embedding was assigned based only on the amino-acid name, so two same amino-acids with different number of hydrogen atoms will have exactly the same embeddings. Interestingly, the latter gave better results in most of the cases. Furthermore, the edge attributes were not considered and the prediction was done solely on one node attribute.

The core SchNet architecture processes these nodes through multiple interaction layers, where the representation of each node is shaped by its local environment. The continuous-filter convolution employed in the interaction layers help encode spatial information faithfully in situation — as we have here — where the nodes are not placed on a uniform grid. In our model we used 5 interaction layers, after which the SchNet representation is passed through a Multi-Layer Perceptron (MLP) featuring a \texttt{Linear} layer, \texttt{shifted soft-plus} activation, and a final \texttt{Linear} layer. 

\paragraph{Training and Specialised Loss Functions:} 
Once the desired resolution of the histogram was fixed, along with the histogram boundaries (corresponding to conservative estimates of the extremal values of the predicted spectra — typically -2 and 2), the parameters of the SchNet such as the length of the embeddings, the size of the hidden layers, the number of interaction layers etc. was fixed using a hyperparameter optimization process. 

Given a choice of the parameters, the training of the model was performed using the Adam optimizer with batched data loaders. A notable feature of our training framework is the flexibility in loss definitions. While the default criterion is the $L_1$ Loss (Mean Absolute Error), the pipeline supports specialized soft-constraint penalty formulations. When predicting eigenspectrum histograms, the total number of physical eigenvalues (derived from the total number of electrons) is a known constant. To enforce this physical invariant, a penalty term $\lambda (\sum Y_{pred} - \sum Y_{true})^2$ can be added to the Mean Squared Error to strictly discourage non-physical predictions. Furthermore, to prevent mode collapse, the training loop continuously monitors the batch variance of predictions, automatically halting execution if the model becomes insensitive to diverse structural inputs.

\subsection{Random Forests and Radial Distribution Functions}

\paragraph{Data Preparation and Scaling:}
The Random Forest inputs were constructed utilizing unweighted and weighted Radial Distribution Functions extracted into tabular formats. The dataset underwent an 80/20 train/test split. To standardize these structural descriptors prior to training, multiple data scaling methodologies were systematically evaluated within the pipeline, including \texttt{RobustScaler}, \texttt{MaxAbsScaler}, \texttt{MinMaxScaler}, and \texttt{StandardScaler}. To thoroughly assess the efficacy of elemental weighting, executions dynamically evaluated the RDF features modulated by distinct physical factors---namely covalent radii, van der Waals (vdW) radii, and valence shell electrons.

\paragraph{Model Architecture and Hyperparameter Tuning:}


Hyperparameter optimization of the Random Forest model was performed using exhaustive grid search combined with 3-fold cross-validation, minimizing the mean squared error on the training set. The following hyperparameters were included in the search: number of estimators $n_{\text{estimators}} \in \{2, 4, 8, 16\}$, maximum tree depth $d_{\text{max}} \in \{\text{None}, 3\}$, fraction of features considered at each split $m_{\text{features}} \in \{0.2, 0.5, 0.8\}$, minimum fraction of samples required to split an internal node $m_{\text{split}} \in \{0.2, 0.5, 0.8\}$, and minimum number of samples required at a leaf node $m_{\text{leaf}} \in \{2, 4, 8\}$. Since the prediction target was multivariate, the Random Forest was wrapped in a \texttt{MultiOutputRegressor}, fitting one independent estimator per output. The optimal hyperparameter combination was selected based on the best mean cross-validation score across all folds.

\subsection{Validation, Results, and Comparative Analysis}

\begin{figure}[htbp]
    \centering
    \definecolor{tabblue}{HTML}{1F77B4}
    \definecolor{taborange}{HTML}{FF7F0E}
    \definecolor{tabgreen}{HTML}{2CA02C}

    \renewcommand{\arraystretch}{2} 

    \begin{tabular}{c c c}


        \begin{tikzpicture}[baseline]
            \begin{axis}[
                width=0.33\linewidth, height=4.5cm,
                title={{\large (a)} System: 1P4B, KRR},
                xlabel={Eigenvalue Index},
                ylabel={Eigenvalue},
                title style={font=\bfseries},
                legend pos=south east,
                legend cell align={left},
                legend style={draw=black!30, fill=white, font=\tiny},
                ymin=-1.3, ymax=0.1,
                xmin=-200, xmax=5100,
            ]
            \addplot[color=tabblue, line width=1pt] table [y index=0, x expr=\coordindex] {attachments/KRR_data_1P4B_true.txt};
            \addlegendentry{eigenvalues true}
            \addplot[color=taborange, line width=1pt] table [y index=0, x expr=\coordindex] {attachments/KRR_data_1P4B_pred.txt};
            \addlegendentry{eigenvalues predicted}
            \addplot[color=tabgreen, line width=1pt] table [y index=0, x expr=\coordindex] {attachments/KRR_data_1P4B_pred_shifted.txt};
            \addlegendentry{eigenvalues predicted shifted}
            \end{axis}
        \end{tikzpicture} &

        \begin{tikzpicture}[baseline]
            \begin{axis}[
                width=0.33\linewidth, height=4.5cm,
                title={System: 1P4B, GNN},
                yticklabels={,,},
                xlabel={Eigenvalue Index},
                title style={font=\bfseries},
                legend pos=south east,
                legend cell align={left},
                legend style={draw=black!30, fill=white, font=\tiny},
                ymin=-1.3, ymax=0.1,
                xmin=-200, xmax=5100,
            ]
            \addplot[color=tabblue, line width=1pt] table [y index=0, x expr=\coordindex] {attachments/GNN_data_1P4B_true.txt};
            \addlegendentry{eigenvalues true}
            \addplot[color=taborange, line width=1pt] table [y index=0, x expr=\coordindex] {attachments/GNN_data_1P4B_pred.txt};
            \addlegendentry{eigenvalues predicted}
            \addplot[color=tabgreen, line width=1pt] table [y index=0, x expr=\coordindex] {attachments/GNN_data_1P4B_pred_shifted.txt};
            \addlegendentry{eigenvalues predicted shifted}
            \end{axis}
        \end{tikzpicture} &

        \begin{tikzpicture}[baseline]
            \begin{axis}[
                width=0.33\linewidth, height=4.5cm,
                title={System: 1P4B, RF},
                yticklabels={,,},
                xlabel={Eigenvalue Index},
                title style={font=\bfseries},
                legend pos=south east,
                legend cell align={left},
                legend style={draw=black!30, fill=white, font=\tiny},
                ymin=-1.3, ymax=0.1,
                xmin=-200, xmax=5100,
            ]
            \addplot[color=tabblue, line width=1pt] table [y index=0, x expr=\coordindex] {attachments/RF_data_1P4B_true.txt};
            \addlegendentry{eigenvalues true}
            \addplot[color=taborange, line width=1pt] table [y index=0, x expr=\coordindex] {attachments/RF_data_1P4B_pred.txt};
            \addlegendentry{eigenvalues predicted}
            \addplot[color=tabgreen, line width=1pt] table [y index=0, x expr=\coordindex] {attachments/RF_data_1P4B_pred_shifted.txt};
            \addlegendentry{eigenvalues predicted shifted}
            \end{axis}
        \end{tikzpicture} \\


        \begin{tikzpicture}[baseline]
            \begin{axis}[
                width=0.33\linewidth, height=4.5cm,
                title={{\large (b)} System: 1ZSG, KRR},
                xlabel={Eigenvalue Index},
                ylabel={Eigenvalue},
                title style={font=\bfseries},
                legend pos=south east,
                legend cell align={left},
                legend style={draw=black!30, fill=white, font=\tiny},
                ymin=-1.3, ymax=0.1,
                xmin=-100, xmax=2050,
            ]
            \addplot[color=tabblue, line width=1pt] table [y index=0, x expr=\coordindex] {attachments/KRR_data_1ZSG_true.txt};
            \addlegendentry{eigenvalues true}
            \addplot[color=taborange, line width=1pt] table [y index=0, x expr=\coordindex] {attachments/KRR_data_1ZSG_pred.txt};
            \addlegendentry{eigenvalues predicted}
            \addplot[color=tabgreen, line width=1pt] table [y index=0, x expr=\coordindex] {attachments/KRR_data_1ZSG_pred_shifted.txt};
            \addlegendentry{eigenvalues predicted shifted}
            \end{axis}
        \end{tikzpicture} &

        \begin{tikzpicture}[baseline]
            \begin{axis}[
                width=0.33\linewidth, height=4.5cm,
                title={System: 1ZSG, GNN},
                yticklabels={,,},
                xlabel={Eigenvalue Index},
                title style={font=\bfseries},
                legend pos=south east,
                legend cell align={left},
                legend style={draw=black!30, fill=white, font=\tiny},
                ymin=-1.3, ymax=0.1,
                xmin=-100, xmax=2050,
            ]
            \addplot[color=tabblue, line width=1pt] table [y index=0, x expr=\coordindex] {attachments/GNN_data_1ZSG_true.txt};
            \addlegendentry{eigenvalues true}
            \addplot[color=taborange, line width=1pt] table [y index=0, x expr=\coordindex] {attachments/GNN_data_1ZSG_pred.txt};
            \addlegendentry{eigenvalues predicted}
            \addplot[color=tabgreen, line width=1pt] table [y index=0, x expr=\coordindex] {attachments/GNN_data_1ZSG_pred_shifted.txt};
            \addlegendentry{eigenvalues predicted shifted}
            \end{axis}
        \end{tikzpicture} &

        \begin{tikzpicture}[baseline]
            \begin{axis}[
                width=0.33\linewidth, height=4.5cm,
                title={System: 1ZSG, RF},
                yticklabels={,,},
                xlabel={Eigenvalue Index},
                title style={font=\bfseries},
                legend pos=south east,
                legend cell align={left},
                legend style={draw=black!30, fill=white, font=\tiny},
                ymin=-1.3, ymax=0.1,
                xmin=-100, xmax=2050,
            ]
            \addplot[color=tabblue, line width=1pt] table [y index=0, x expr=\coordindex] {attachments/RF_data_1ZSG_true.txt};
            \addlegendentry{eigenvalues true}
            \addplot[color=taborange, line width=1pt] table [y index=0, x expr=\coordindex] {attachments/RF_data_1ZSG_pred.txt};
            \addlegendentry{eigenvalues predicted}
            \addplot[color=tabgreen, line width=1pt] table [y index=0, x expr=\coordindex] {attachments/RF_data_1ZSG_pred_shifted.txt};
            \addlegendentry{eigenvalues predicted shifted}
            \end{axis}
        \end{tikzpicture} 

    \end{tabular}

    \caption{Exact interpolated reconstructed eigenvalues across two diverse protein complexes (rows), plotted directly from numerical outputs, using three different machine learning predictors (columns): KRR, GNN, and RF. (a) 1P4B system (4,901 eigenvalues), sharing a common x-axis range of $(-200, 5100)$. (b) 1ZSG system (1,947 eigenvalues), sharing a common x-axis range of $(-100, 2050)$. All six plots utilize a consistent y-axis range of $(-1.3, 0.1)$. Note how the applied global constant shift aligns the predicted lowest energy state (green) perfectly with the exact baseline (blue) in all models and systems.}
    \label{fig:combined_exact_predictions}
\end{figure}
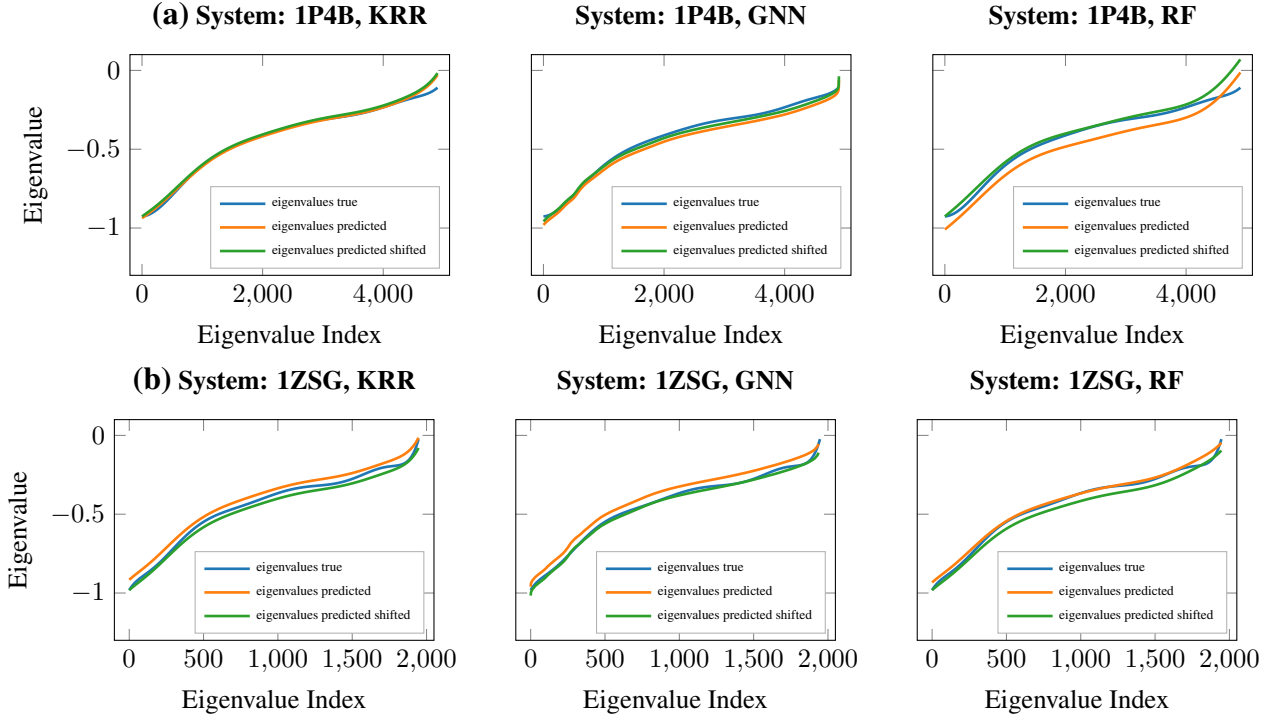

\paragraph{Kernel Ridge Regression Results:} Following optimization, the best KRR estimator was deployed against a hold-out test set. Global predictive quality was assessed through standard metrics, including MAE, MSE, and the overarching $R^2$ score. Furthermore, continuous interpolations of the discrete Chebyshev predictions were reconstructed over a normalized linear space. As shown in Fig. \ref{fig:combined_exact_predictions}, a consistent characteristic emerged across the generated models: the predicted continuous spectrum (orange line) generally maintained an accurate structural shape but exhibited a uniform global displacement (shift) relative to the true spectrum (blue line). To systematically cure this, our post-processing routine computed the difference between the actual lowest eigenvalue, computed to relatively low accuracy, and the predicted lowest eigenvalue, applying their difference to shift the entire predicted vector back into strict alignment (green line).



\paragraph{Graph Neural Network Results:} Following the prediction, outputs tracking histogram distributions are post-processed via a strict sum-preserving rounding algorithm. This algorithm rounds fractional bin predictions down to integers and systematically redistributes the remainder back to the bins with the highest fractional deficits to perfectly match the known total number of physical eigenvalues. The predicted eigenspectrum of the systems 1P4B and 1ZSG is shown in Fig.~\ref{fig:combined_exact_predictions}. Interestingly, the node attributes based only on amino-acid names were enough to obtain a very good prediction of the eigenspectrum. However, in both systems, the prediction towards the edges starts to diverge from the real eigenvalues due to falls prediction of the number of eigenvalues in the bins closer to the edge. This was partially alleviated by applying the mask, a simpler prediction if a bin should be empty. The model with the mask applied clearly demonstrates that even simple attributes, like amino-acid names, were enough for GNN to grasp the system properties.


Similar as the KRR model, the resulting model was also evaluated across the test subset using an array of regression metrics: Mean Squared Error (MSE), Mean Absolute Error (MAE), $R^2$ Score, and Explained Variance. Identical to the KRR workflow, continuous spectrum reconstruction necessitated a post-prediction shift. By calibrating the interpolated prediction curve to match the exact lowest physical eigenvalue, the Random Forest model consistently produced structurally and numerically accurate continuous eigenspectra.



\paragraph{Random Forest Results:} Similar as in the KRR model, Random Forest models were trained to predict the Chebyshev polynomial expansion coefficients representing the molecular eigenspectrum, using RDF feature vectors weighted by five different atomic properties and normalized using five distinct scaling strategies. Model performance was evaluated using MAE, MSE, and the coefficient of determination ($R^2$), each reported as the minimum, maximum, and mean across the cross-validation folds. All weighting variants achieved consistently high predictive accuracy, with mean $R^2$ values ranging from 0.951 to 0.963 across all configurations. The electronegativity-weighted RDF without feature scaling yielded the highest mean $R^2$ of 0.963, accompanied by a mean MAE of $4.39\times 10^{-3}$ and a mean MSE of $2.40\times 10^{-4}$. The covalent radius weighting also performed competitively, achieving mean $R^2$ values between 0.957 and 0.960 across scaling strategies, with the no-scaling variant attaining a mean $R^2$ of 0.959. 

Interestingly, the unweighted RDF variant ($A_i=A_j=1$) produced results broadly comparable to the weighted variants, maintaining mean $R^2$ values between 0.954 and 0.962. This suggests that even a purely geometric descriptor captures substantial information regarding the eigenspectrum. Among the remaining variants, valence electron and van der Waals (vdW) radius weightings showed slightly lower mean $R^2$ values, with the vdW variant under \texttt{MaxAbsScaler} yielding the lowest mean $R^2$ of 0.947 observed across all configurations.

The choice of feature scaling had a modest but consistent effect on model performance. Across all atomic weighting schemes, the no-scaling baseline and \texttt{RobustScaler} tended to perform on par with, or slightly above, the remaining scaling strategies. \texttt{MinMaxScaler} and \texttt{StandardScaler} occasionally introduced marginal degradations in the mean $R^2$, though no single scaling method was uniformly superior across all weighting variants. The narrow spread of mean $R^2$ values across the scaling strategies (typically within 0.01) strongly indicates that the models are robust to the choice of input normalization.

Overall, maximum $R^2$ values approaching or exceeding 0.999 were recorded for all configurations, demonstrating that individual cross-validation folds can achieve near-perfect reconstructions of the Chebyshev coefficients. Similar to the KRR workflow, continuous spectrum reconstruction necessitated a post-prediction shift. By calibrating the interpolated prediction curve to match the exact lowest physical eigenvalue, the Random Forest model consistently produced structurally and numerically accurate continuous eigenspectra.

\begin{figure}[htbp]
    \centering
    \includegraphics[width=\textwidth]{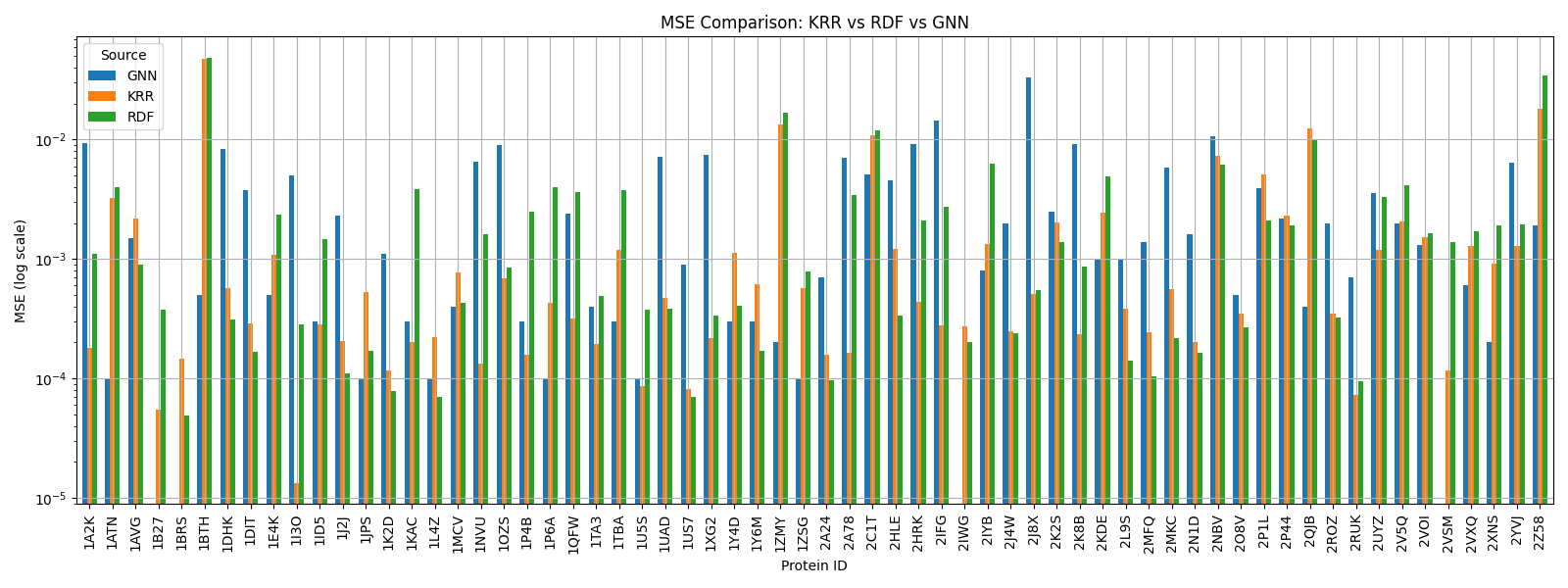}
    \caption{MSE analysis of KRR, RF and GNN frameworks compared to the real eigenvalues for all 64 systems in test set.}
    \label{fig:MSE_comparison}
\end{figure}

\paragraph{Comparative Summary and Discussion:}
In summary, all three machine learning frameworks demonstrate a robust capability to accurately capture and reconstruct the complex features of the molecular eigenspectrum, proving that data-driven models are entirely viable for bypassing explicit, costly Hamiltonian diagonalizations. Remarkably, simpler classical approaches such as KRR and Random Forests perform exceptionally well, in majority test systems achieving smaller MSE than GNN (see Figure~\ref{fig:MSE_comparison}. Despite their lower architectural complexity, these simpler models deliver predictive accuracies that are highly competitive with, and frequently comparable to, the significantly more sophisticated and computationally expensive GNN. This indicates that well-chosen structural features --- such as randomized spectral snapshots or radial distribution functions --- encapsulate sufficient physical information to allow classical methods to thrive. Ultimately, while the GNN offers unique benefits in directly mapping graph topologies to density-of-states histograms, the strong performance of KRR and Random Forest underscores that lighter, less intensive models provide a highly effective and cost-efficient alternative for these physical systems.

\subsection{Integration within the Self-Consistent Cycle and Acceleration Analysis}
\label{sec:scf_integration}

The ultimate validation of data-driven spectral predictors lies in their practical capacity to accelerate the Self-Consistent Field (SCF) cycle. In Kohn-Sham Density Functional Theory (KS-DFT), the computational bottleneck is dictated by the iterative solution of the single-particle eigenvalue problem:
\begin{equation}
    \hat{H}[\rho^{(n)}] \psi_i^{(n)} = \epsilon_i^{(n)} \psi_i^{(n)},
\end{equation}
where $\rho^{(n)}$ is the electronic density at iteration $n$, and $\epsilon_i^{(n)}$ represents the corresponding energy eigenvalues. To evaluate the efficacy of our machine learning (ML) models, we project the predicted spectral density directly against the explicit trajectory of a standard BigDFT simulation for the 1ZSG protein system (comprising 1,861 atoms).

\begin{figure}[htbp]
    \centering
    \includegraphics[width=0.45\textwidth]{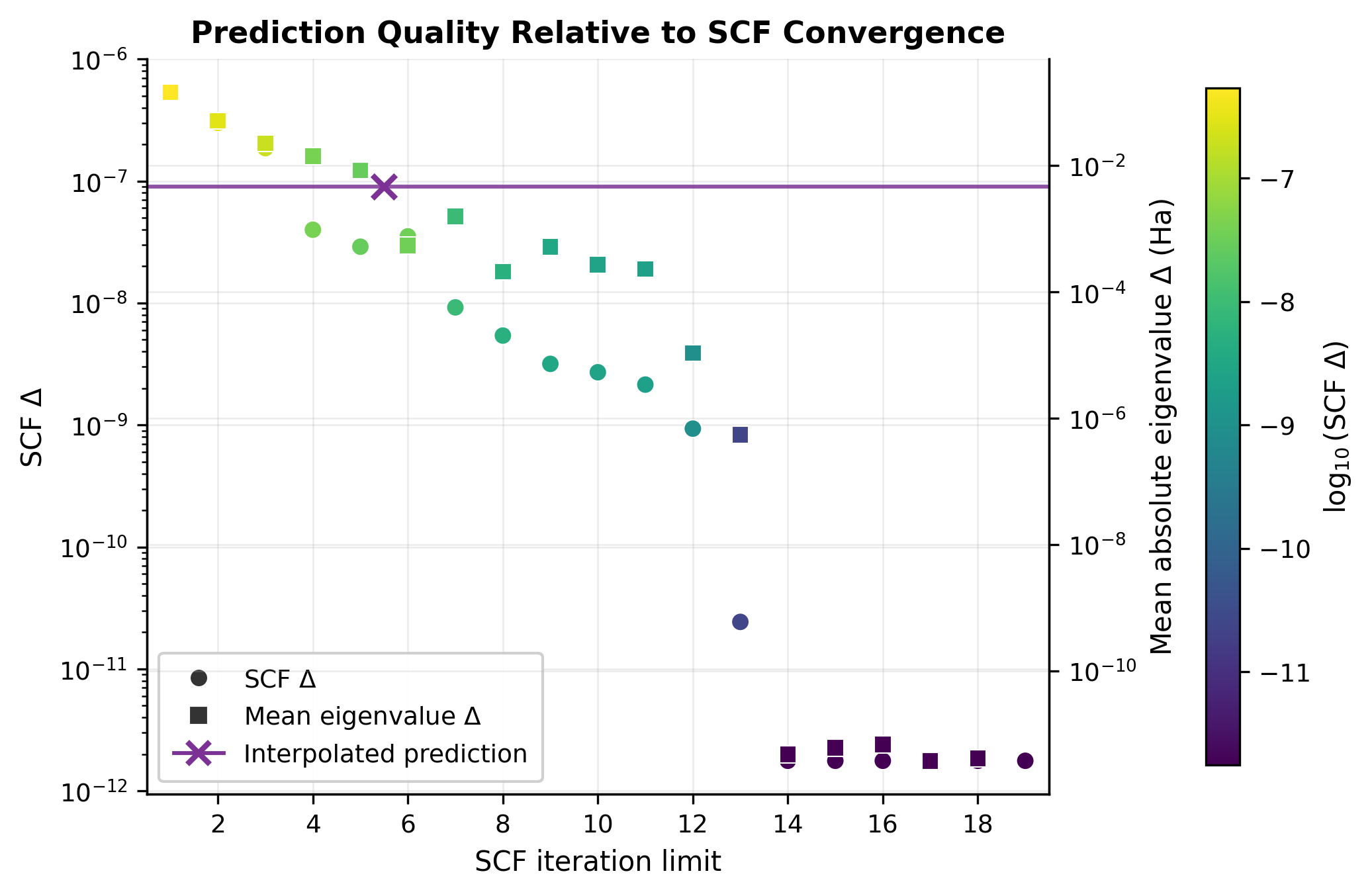}
    \includegraphics[width=0.45\textwidth]{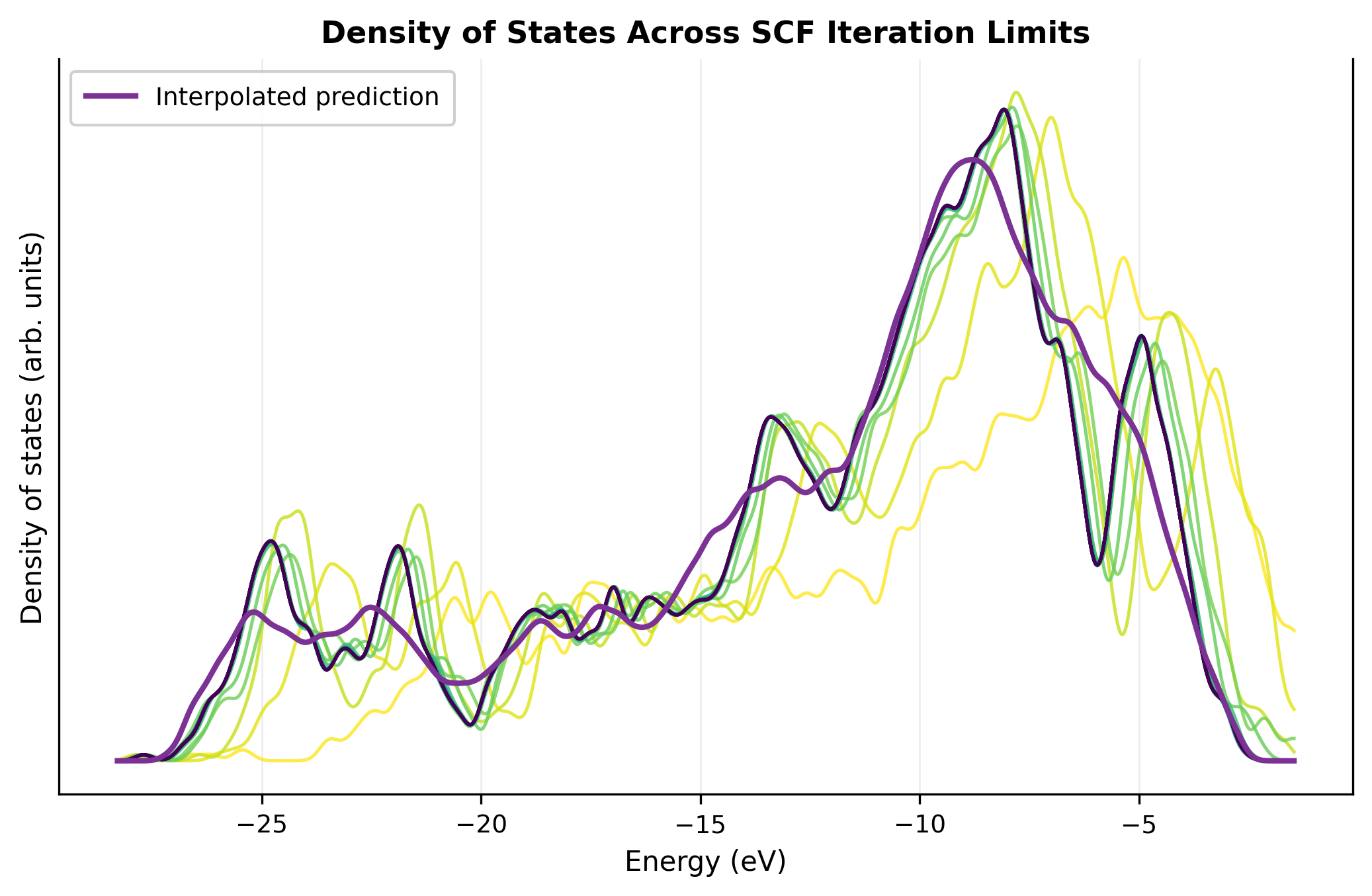}
    \caption{Quantitative analysis of spectral prediction quality on the 1ZSG system (1,861 atoms). Left: The predicted spectral precision (purple cross) mapped against the SCF convergence metrics. The left axis denotes the density residual ($\text{SCF }\Delta$), while the right axis measures the mean absolute error of the eigenvalue spectrum ($\text{Avg Eigenvalue }\Delta$). Right: Dynamic evolution of the Density of States (DoS) across successive SCF iterations (color-coded from yellow to dark blue) contrasted against the ML-predicted spectral profile (solid purple line).}
    \label{fig:scf_acceleration}
\end{figure}

We quantify the convergence state using two concurrent metrics: the density-based residual $\text{SCF }\Delta^{(n)} = \int |\rho^{(n)}(\mathbf{r}) - \rho^{(n-1)}(\mathbf{r})|\, d\mathbf{r}$, and the mean spectral error relative to the fully converged ground state eigenvalues $\epsilon_i^{(\infty)}$:
\begin{equation}
    \text{Avg Eigenvalue }\Delta^{(n)} = \frac{1}{N_{\text{states}}} \sum_{i=1}^{N_{\text{states}}} \left| \epsilon_i^{(n)} - \epsilon_i^{(\infty)} \right|.
\end{equation}

As illustrated in Figure \ref{fig:scf_acceleration} (left), the spectral precision achieved by our Random Forest model serves as an exceptionally high-quality initial ansatz. The ML-predicted eigenvalues exhibit a mean error functionally equivalent to an SCF convergence threshold of $\text{SCF }\Delta \approx 10^{-7}$. Mathematically, this maps directly to the intersection point at $n \approx 5.5$. This indicates that the data-driven model effectively bypasses the initial, highly non-linear phase of the density optimization, effectively sparing the overhead of approximately 5 to 6 full self-consistent iterations.

A deeper look into the spectral features is provided by the Density of States (DoS) evolution in Figure \ref{fig:scf_acceleration} (right). Early-stage SCF iterations (light yellow profiles) fail to capture the fine quantum mechanical features, exhibiting smoothed, unphysical peak merges particularly in the deep valence band between $-25$~eV and $-20$~eV, as well as shifts near the Fermi level ($\sim -5$~eV). Conversely, the ML model (purple line) reconstructs these intricate electronic structures with high fidelity, replicating the peak splitting and local density variations of the fully converged Hamiltonian. 



 \section{Conclusions and Future Work}

  The LimitX project demonstrates that data-driven spectral prediction is a viable pathway for reducing the cost of large-scale electronic-structure workflows. By combining physically restricted training domains, fragment-based molecular representations, and compact spectral targets based on Chebyshev coefficients or density-of-states, we
  transform a variable-size eigenspectrum prediction problem into a tractable machine-learning task. Across KRR, GNN, and RF models, the predicted spectra reproduce the main features of the reference eigenspectra after a simple global alignment procedure.

  The SCF analysis on the 1ZSG system provides a first practical interpretation of this prediction quality. The predicted spectrum reaches an average eigenvalue error comparable to that obtained after approximately 5--6 standard SCF iterations. This indicates that the ML model captures spectral information characteristic of an already partially converged   calculation and may therefore help bypass the most unstable early phase of the self-consistent cycle. However, this result should be understood as an a posteriori quality assessment rather than a complete end-to-end acceleration benchmark. A series of direct BigDFT calculation initialized with ML-predicted spectral or kernel information would be an ideal futher step to quantify the actual reduction in SCF iterations, wall time, and solver cost.

  A closely related next step is to perform the same type of validation for fragment-level quantities, in particular fragment bond orders derived from the density matrix reduced on the fragment. Since bond orders enter the chemical interpretation and fragmentation quality of the system, predicting them accurately would provide an additional route for accelerating or stabilizing the construction of chemically
  meaningful reduced representations. Establishing whether bond-order predictions exhibit convergence quality comparable to intermediate SCF states would make the present spectral analysis more complete.

  Looking forward, the predicted spectral information should be integrated directly into numerical eigensolver workflows, especially the FrASE rational filter-based eigensolver currently under development. In this context, ML predictions can guide subspace allocation, estimate density-of-states structure, improve filter placement, and support adaptive
  solver choices. The key remaining task is therefore to move from prediction-quality diagnostics to operational benchmarks where the ML-assisted workflow is compared directly against standard BigDFT and eigensolver runs.
In particular, the predicted spectra are expected to support:
  \begin{itemize}
      \item \textbf{Precise Subspace Allocation:} estimating the Density of States (DOS) to identify clustered and sparse spectral regions, enabling better allocation of subspace sizes and reducing over-provisioning.
      \item \textbf{Krylov Subspace Reuse:} Leveraging the predicted spectral data to construct robust polynomial preconditioners, which will accelerate linear solves across multiple shifted systems simultaneously within a shared Krylov subspace.
      \item \textbf{Adaptive Solver Selection:} using spectral diagnostics as a recommender signal to choose between robust block methods and lighter iterative strategies depending on local spectral structure.
  \end{itemize}



\begin{acknowledgement}
  This project (LimitX) has received funding from the European High-Performance Computing Joint Undertaking (JU) under grant agreement No. 101118139. The JU receives support from the European Union's Horizon Europe Programme.
\end{acknowledgement}

\vspace{\baselineskip}

\end{document}